# Strain-modulated Bandgap and Piezo-resistive Effect in Black Phosphorus Field-effect Transistors

Zuocheng Zhang[1,2†], Likai Li[1,2,3†], Jason Horng[3†], Nai Zhou Wang[4,5,6], Fangyuan Yang[1,2], Yijun Yu[1,2], Yu Zhang[1,2], Guorui Chen[1,2], Kenji Watanabe[7], Takashi Taniguchi[7], Xian Hui Chen[4,5,6*], Feng Wang[3,8,9*] and Yuanbo Zhang[1,2,6*]

[1]*State Key Laboratory of Surface Physics and Department of Physics, Fudan University, Shanghai 200433, China.*

[2]*Institute for Nanoelectronic Devices and Quantum Computing, Fudan University, Shanghai 200433, China.*

[3]*Department of Physics, University of California at Berkeley, Berkeley, California 94720, USA.*

[4]*Hefei National Laboratory for Physical Science at Microscale and Department of Physics, University of Science and Technology of China, Hefei, Anhui 230026, China.*

[5]*Key Laboratory of Strongly Coupled Quantum Matter Physics, University of Science and Technology of China, Hefei, Anhui 230026, China.*

[6]*Collaborative Innovation Center of Advanced Microstructures, Nanjing 210093, China.*

[7]*National Institute for Materials Science, 1-1 Namiki, Tsukuba, 305-0044, Japan*

[8]*Materials Sciences Division, Lawrence Berkeley National Laboratory, Berkeley, California 94720, USA.*

[9]*Kavli Energy NanoSciences Institute at the University of California, Berkeley and the Lawrence Berkeley National Laboratory, Berkeley, California, 94720, USA.*

† These authors contributed equally to this work.

* Email: zhyb@fudan.edu.cn; fengwang76@berkeley.edu; chenxh@ustc.edu.cn

**Energy bandgap largely determines the optical and electronic properties of a semiconductor. Variable bandgap therefore makes versatile functionality possible in a single material. In layered material black phosphorus[1–5], the bandgap can be modulated by the number of layers; as a result, few-layer black phosphorus has discrete bandgap values that are relevant for opto-electronic applications in the spectral range from red, in monolayer, to mid-infrared in the bulk limit[3,6–8]. Here, we further demonstrate continuous bandgap modulation by mechanical strain applied through flexible substrates. The strain-modulated bandgap significantly alters the charge transport in black phosphorus at room temperature; we for the first time observe a large piezo-resistive effect in black phosphorus field-effect transistors (FETs). The effect opens up opportunities for future development of electro-mechanical transducers based on black phosphorus, and we demonstrate strain gauges constructed from black phosphorus thin crystals.**

Under mechanical strain, the deformation of the atomic lattice is able to induce profound changes in the electronic structure of a crystalline material. This is best exemplified in doped silicon, where strain alters the energy bands of electron or hole carriers; the transfer of carriers to bands with small effective mass leads to drastic enhancement of carrier mobility (and thus conductivity)[9–13]. Strained silicon is, therefore, able to provide improved switching performance as the transistor dimension is aggressively scaled down in modern electronics[14,15]. Meanwhile, the large resistance response under strain has enabled the development of a myriad of silicon-based transducers[16], such as strain and torque gauges, that are widely used in industrial applications.

Black phosphorus, a two-dimensional (2D) material with a puckered honeycomb lattice, offers new possibilities. The puckered lattice in a monolayer, shown in Fig. 1a, can be viewed as rows of two orthogonally coupled hinges along the zigzag direction[17]. The structure makes black phosphorus a soft, and yet mechanically resilient, material that can withstand large strain modulation[18]. More importantly, deformation of the puckers under strain changes the configuration of $p_z$ orbitals near the band edges[19]; modulating the black phosphorus bandgap (and therefore its material properties) via strain becomes a possibility. Indeed, it has been shown that a moderate high pressure of ~1.2 GPa can close the bandgap, and raise the conductance by one order of magnitude at room temperature[20]; modulation of the optical properties was also observed in corrugated black phosphorus sheets[21]. Those studies hinted at strain as a powerful tool to modulate the electronic and optical properties of black phosphorus.

Here, we observe a large piezo-resistive effect in black phosphorus FETs at room temperature. The piezo-resistive response (defined as the relative change in sample resistance, $\delta R/R$, at a given strain, $\varepsilon$) varies with the gate doping, and reaches maximum at the charge neutral point (CNP). We attribute the large room-temperature piezo-resistance to the strain-modulated bandgap, which we directly observe in the infrared (IR) spectroscopic measurements on strained black phosphorus. Our results highlight two distinct advantages of black phosphorus compared to other electro-mechanical materials. First, the direct bandgap can be continuously tuned by strain; in few-layer specimens, such tunability may provide black phosphorus continuous coverage of a technologically important spectral range from far-infrared to visible. Second, the large piezo-resistance, combined with black phosphorus's mechanical robustness under large strain[18], may lead to sensitive transducers with large dynamic range that are useful in flexible electronics and other industrial applications. To this end, we demonstrate a strain sensor constructed from black phosphorus flakes that operates under ambient condition.

We use flexible film as substrate of black phosphorus FET, and induce strain in the device by bending the substrate. Fig. 1b displays an optical image of a typical device, and the cross-sectional view of the device is schematically shown in Fig. 1c. We constructed the device using the dry-transfer technique described in ref. 22. We started with a black phosphorus flake cleaved onto a polypropylene carbon (PPC) film, and used the flake on the film to pick up a hexagonal boron nitride (hBN) flake from a $SiO_2$/Si wafer surface. The black phosphorus crystals were typically ~20 nm thick, and hBN flakes are ~8-30 nm thick, as estimated from their optical contrast. The black phosphorus/hBN stack was subsequently released onto a graphite flake (which later serves as the back gate) supported on a flexible polyethylene naphthalate (PEN) substrate. We then defined electrodes (Cr/Au with thicknesses of 5 nm and 45 nm, respectively) on black phosphorus using standard electron beam lithography; the electrodes also serve as clamping points to prevent sample slippage under mechanical strain. Finally, a second layer of hBN was transferred on top to help protect the sample from degradation during measurements in air. (We further used a drop of silicone vacuum grease placed on the device for additional protection in air.) To further mitigate sample degradation, all fabrication processes, except for the electron beam lithography, were performed in an argon atmosphere with $H_2O$ and $O_2$ content kept below 1 ppm.

Bending the flexible PEN substrate upward/downward induces a continuously variable, uniaxial strain in black phosphorus FET (Fig. 1c). In a continuum mechanics model[23] for elastic film, the strain at the surface of the substrate is determined by substrate thickness, $h$, and the radius of the curvature, $r$; specifically, $\varepsilon = h/2r$. Here the thickness of the PEN film is $h = 125$ μm, and we extract $r$ from the profile of the bent substrate (Supplementary Information). If no slippage occurs in the FET heterostructure assembly, such $\varepsilon$ is directly transferred to the black phosphorus crystal; a positive/negative $\varepsilon$ denotes a tensile/compressive strain. Unless specified otherwise, we monitor the longitudinal sample resistance, $R$, in a four-terminal setup to eliminate spurious signal from electrical contacts as the strain is varied. Meanwhile, a gate voltage, $V_g$, applied on the graphite back gate modulates the charge carrier density in the FET. All measurements were performed at room temperature.

We observe a large piezo-resistive effect at the CNP of the black phosphorus FET. At zero strain, $R$ measured as a function of $V_g$ exhibits ambipolar behaviour, and the CNP manifests as the peak in $R$ (black curve, Fig. 1d). The fact that CNP appears at $V_g > 0$ indicates that the sample is hole-doped, possibly from reaction with $H_2O$ and $O_2$ in the environment[24]. A tensile/compressive strain does not change the position of the CNP; it, however, dramatically increases/suppresses the peak value of $R$ at the CNP. As shown in Fig. 1d, a large piezo-resistive response up to $\delta R/R \approx 25\%$ was recorded for a moderate tensile strain of $\varepsilon = 0.15\%$ at the CNP (red curve), and the response reached $-15\%$ for a compressive strain of $\varepsilon = -0.18\%$ (blue curve). The piezo-resistive response is maximum at the CNP, and diminishes as $V_g$ moves away from the CNP.

We attribute the large piezo-resistive effect at the CNP to strain-modulated bandgap in black phosphorus. At the CNP, the free charge carriers near the bottom black phosphorus/hBN are depleted by the gate, so the charge transport is dominated by thermally excited electrons and holes. Indeed, $R$ measured at the CNP increases exponentially as the temperature, $T$, is lowered. Analysis of the thermally activated behaviour indicates that excitations across the bandgap are responsible for the charge transport at room temperature and above (Supplementary Information). Strain can, therefore, effect significant change in $R$ through the modulation of the bandgap. Such bandgap modulation was seen in corrugated black phosphorus flake[21], but a systematic study is needed to quantitatively determine the strain dependence.

To this end, we performed IR spectroscopy on black phosphorus under a variable strain to quantitatively delineate the strain-modulated bandgap. We used devices with a simplified structure for spectroscopic measurements: two Au contacts anchor the black phosphorus flake to the PEN substrate; a layer of hBN is placed on top for protection (Fig. 2a). Prior to introducing strain in the flake, we determined the orientation of the crystal using the strong polarization dependence of the IR spectra (Fig. 2b; see also ref. 8. The crystal orientation is indicated by arrows in Fig. 2a). We found that mechanical strain, induced either along armchair or zigzag direction, significantly alters the size of bandgap, $E_g$; the effect is manifested as appreciable blueshift/redshift of the absorption peak in the reflection spectra under tensile/compressive strains (Fig. 2c and 2d). (Here the reflection contrast is defined as $1 - r_s/r_{ref}$, where $r_s$ and $r_{ref}$ is the reflectance of the sample and Au film, respectively. We induce tensile/compressive strain by conforming the flexible PEN substrate to convex/concave surfaces with known curvatures.) The observed trend agrees with previous calculations[19,25–29].

More importantly, the data set enables us to quantitatively extract $E_g$ as a function of strain induced along both armchair and zigzag directions (Fig. 2e; we neglected the small exciton contribution in extracting $E_g$ from the IR spectra[8]). Specifically, $E_g$ is well described by

$$E_g(\varepsilon) = E_0 + \alpha \cdot \varepsilon \tag{1}$$

for strains up to $\varepsilon \approx 0.5\%$, beyond which $E_g(\varepsilon)$ starts to deviate from the linear dependence, possibly because of slippage. Here $E_0 \approx 340$ meV is the bandgap of pristine bulk black phosphorus, and $\alpha$ is the rate of the strain modulation. From line fits to the data, we obtained an $\alpha$ of $99 \pm 4$ meV/% in the armchair direction and $109 \pm 2$ meV/% in the zigzag direction (Fig. 1e). The comparable $\alpha$ values in the two orthogonal directions implies little anisotropy, if at all, in the strain modulation of the bandgap, even though black phosphorus crystal structure is highly anisotropic. Such lack of anisotropy, as well as the value of $\alpha$, is however in excellent agreement with calculations reported in ref. 28. Finally, we note that the isotropic strain modulation is also reflected in the piezo-resistive effect in black phosphorus; strains in zigzag (Fig. 3a) and armchair direction (Fig. 3b) induce equally large responses in $R$ at the CNP. (The small shift of the CNP and decrease of the $R$ peak value are results of slight sample degradation between measurements; see Supplementary Information for detailed discussion.)

We are now poised to quantitatively delineate the origin of the large piezo-resistive effect in black phosphorus. Important insights come from the behaviour of the sample conductance, $\sigma = 1/R$, under strain modulation: tensile/compressive strain induces an almost rigid downward/upward shift of $\sigma$ measured as a function of $V_g$ (Fig. 3c). There are two main points to notice. First, strain modulates the conductance not only at the CNP, but at all doping levels; a large piezo-resistive response is observed at the CNP only because black phosphorus is least conductive there. Second, the field-effect mobility of the carriers, $\mu$, which is given by the slope of the $\sigma(V_g)$ away from the CNP[30], is not affected by strain. Indeed, $\mu$ remains at $\mu_h \sim 300$ $\mathrm{cm^2V^{-1}s^{-1}}$ on the hole side and at $\mu_e \sim 90$ $\mathrm{cm^2V^{-1}s^{-1}}$ on the electron side for all accessible strain values in our study (Fig. 3c, inset). These observations prompt us to view the room-temperature $\sigma$ as the sum of three components, $\sigma = \sigma_{th} + \sigma_{top} + \sigma_{bot}$. Here $\sigma_{th}$ is the contribution from all thermally activated charge carriers, and the rest of the conduction is simplified into two parts: the top layer conduction, $\sigma_{top}$, that is not gated due to electrostatic screening and the bottom layer conduction, $\sigma_{bot}$, that is tuned by the gate. In Drude model, $\sigma_{th} = n_h e\mu_h + n_e e\mu_e$, where $n_h = n_e \propto \exp(-E_g(\varepsilon)/2k_BT)$. Here $k_B$ is the Boltzmann constant and $T = 300$ K is the room temperature. Similarly, $\sigma_{top}$ and $\sigma_{bot}$ are given by $\sigma_{top(bot)} = n_{top(bot)}e\mu_{top(bot)}$, where $n_{top(bot)}$ and $\mu_{top(bot)}$ are concentration and mobility of the free carriers, respectively, in the top (bottom) layer. It's clear that among all three components of $\sigma$ only $\sigma_{th}$ depends on strain. At the CNP, the gate doping makes $n_{bot} = 0$, so $\sigma = \sigma_{th} + \sigma_{top}$, of which only the former, $\sigma_{th}$, depends on the strain.

Combined with the empirical relation between bandgap and strain obtained in Eq. (1), our analysis leads to a quantitative description of the piezo-resistive response at the CNP:

$$\frac{\delta R}{R} = \frac{1}{1+f}\frac{\delta\sigma_{th}}{\sigma_{th}} \cong \frac{1}{1+f}\frac{\alpha}{2k_BT}\varepsilon \tag{2}$$

where $f = \sigma_{top}/\sigma_{th}$. The gauge factor (GF) of piezo-resistance, a widely used metric defined as the piezo-resistive response to strain, is therefore obtained as the prefactor preceding $\varepsilon$ in Eq. (2). In the limit of vanishing residue parallel conduction in the top layer, i.e. $\sigma_{top} \to 0$, GF reaches a theoretical maximum value of $\alpha/(2k_BT) \sim 220$. In the presence of finite $\sigma_{top}$, however, the GF is reduced by the factor $1/(1+f)$. Indeed, line fits of $\delta R/R$ as a function of $\varepsilon$ yields a GF of ~120 in the black phosphorus FET for strain induced in both armchair and zigzag directions (Fig. 3d). Such a GF is still two orders of magnitude higher than that in the commercial metal film strain gauges, and rivals the highest values obtained in semiconductor strain gauges. Our analysis also indicates that improvement up to the theoretical limit of ~220 is possible through minimizing $\sigma_{top}$, either by protecting the sample from environmental doping or by reducing the sample thickness. We further note that finite $\sigma_{bot}$ in the bottom layer further reduces the GF at doping levels away from the CNP ($f$ now becomes $(\sigma_{top} + \sigma_{bot})/\sigma_{th}$), as illustrated in Fig. 4a-c. So the gate tunability in a FET configuration is essential for achieving large piezo-resistive effect in black phosphorus.

Finally, we demonstrate a black phosphorus strain sensor operating under ambient condition. Here we adopt a two-terminal resistance measurement setup without the back gate for simplified sensor fabrication and operation. The GF of the sensor is ~10 without the benefit of a back gate. As shown in Fig. 4d, a $0.12\%$ manually induced strain reliably produces a ~200 Ohm resistance change in repeated operations; robust sensor operation is similarly demonstrated for a $-0.15\%$ strain (Fig. 4e). The robust piezo-resistive response, combined with the excellent stretchability demonstrated elsewhere[18], makes black phosphorus potentially useful in detecting local strain in Micro-Electro-Mechanical Systems (MEMS) and flexible electronics[31–34]. Finally, we note that sample degradation remains a challenge for stable sensor performance in air over long periods of time (Supplementary Information). The main hurdle is that black phosphorus becomes more and more hole doped in air over time, leading to an unstable GF. Better sample protection with $O_2$ and $H_2O$ barrier coating may resolve this issue[18,35,36].

In summary, we observe a large piezo-resistive effect in black phosphorus FETs when the device is gated to the CNP at room temperature. We uncover unambiguous evidence that such piezo-resistive effect originates from strain modulated bandgap of black phosphorus, which we directly probe using IR spectroscopy on black phosphorus flakes under variable strain. The GF of the piezo-resistance reaches 120, and can theoretically approach 220 in thin, pristine crystals. Finally, we demonstrate robust operation of a black phosphorus strain sensor. Sample stability in air remains an issue, but better device packaging should mitigate the problem. The strain modulated bandgap and large room-temperature piezo-resistive effect make black phosphorus a versatile opto-electronic 2D material.

## Methods

The substrate used in this study is 125-μm-thick flexible PEN film. Black phosphorus flakes mechanically exfoliated on the PEN substrate is used for optical measurements. The optical absorption of black phosphorus is performed by reflectance measurements at room temperature. The incident light is first polarized with a broadband calcite polarizer and then focused onto the black phosphorus flakes in a microscopy setup. A spectrometer equipped with both silicon and InGaAs arrayed detectors is used to collect and analyze the reflected light. For transport measurements, additional graphite back gate with a layer of hBN as the gate dielectric is used in the FET structure supported on the PEN substrate. The sample resistance is measured with the standard AC lock-in technique. A current excitation of 10 nA at 17 Hz is used in the lock-in measurements.

**Acknowledgements**

We thank Jie Guan, David Tománek and Liguo Ma for helpful discussions. We also thank Ruiyan Luo and Tao Xue for their help at initial stage of the experiment. Z.Z.,

L.L., F.Y., Y.Y., Y.Z., G.C. and Y.Z. acknowledge support from the National Key Research and Development Program of China (grant no. 2016YFA0300703), NSF of China (grant nos 11527805, 11425415 and 11421404) and the National Basic Research Program of China (973 Program; grant no. 2013CB921902). J.H. and F.W. acknowledge support from National Science Foundation EFRI program (EFMA-1542741). L.L., Y.Z. and F.W. also acknowledge support from Samsung Global Research Outreach (GRO) Program. Z.Z. acknowledges support from NSF of China (grant no. KRH1512533) and China Postdoctoral Science Foundation (grant no. KLH1512079). Part of the sample fabrication was conducted at Fudan Nano-fabrication Lab. N.Z.W. and X.H.C. acknowledge support from the NSF of China (grant no. 11534010), the 'Strategic Priority Research Program' of the Chinese Academy of Sciences (grant no. XDB04040100) and the National Basic Research Program of China (973 Program; grant no. 2012CB922002). K.W. and T.T. acknowledge support from the Elemental Strategy Initiative conducted by the MEXT, Japan and JSPS KAKENHI (grant nos JP26248061, JP15K21722 and JP25106006).

**Author contributions**

Y.Z., F.W., and X.H.C. co-supervised the project. N.Z.W. grew bulk crystal black phosphorus. Z.Z. fabricated the device and carried out transport measurements. L.L. and J.H. obtained and analyzed absorption spectra. L.L., F.Y., Y.Y. and G.C. helped with sample fabrication. Y.Y. and Y.Z helped designing the experimental setup for inducing variable strain. K. W. and T. T. grew hBN crystal. Z.Z. and Y.Z. wrote the paper with input from all authors.

**Figure captions:**

**Figure 1 | Device configuration and strain-dependent ambipolar transport in black phosphorus FET. a,** Crystal structure of monolayer black phosphorus. The armchair and zigzag direction is labelled as $x$ and $y$ direction, respectively. **b,** Optical image of a black phosphorus FET supported on flexible PEN substrate. The crystal orientation of the black phosphorus flake, marked by black arrows, is determined by Raman spectroscopy (Supplementary Information). **c,** Schematic structure of a black phosphorus FET on flexible PEN substrate. Variable compressive/tensile strains are induced in the FET by bending the substrate downward/upward. **d**, Ambipolar transport behaviour of black phosphorus FET under three representative strain levels. Large piezo-resistive response is observed, and the response reaches maximum at the resistance peak, i.e. the CNP. Strains are induced in the zigzag direction.

**Figure 2 | Strain-modulated bandgap in black phosphorus. a,** Optical image of a black phosphorus device on PEN substrate (Sample A) used for IR spectroscopy measurements. The thickness of the black phosphorus flake is approximately 20 nm. Metal contacts anchor the flake to the substrate, and the sample is covered with a layer of hBN for protection. The crystal orientation of the black phosphorus flake, marked by black arrows, was determined by polarization-dependent IR spectra of black phosphorus. **b,** Reflection spectra of black phosphorus under $x$- and $y$-polarized incident light. The anisotropy of the spectra is a result of anisotropic electronic structure of black phosphorus. Under $x$-polarized incident light, the energy position of the spectral peak indicates the size of the bandgap in pristine black phosphorus. **c** and **d,** Evolution of absorption spectra with varying strains induced in armchair (**c**) and zigzag direction (**d**). In both cases, compressive/tensile strain causes redshift/blueshift of the spectral peak. Spectra are shifted by multiples of 6% for clarity. **e,** Bandgap plotted as

a function of strain. Bandgap values were extracted from spectra of two samples: Sample A (thickness ~20 nm) and Sample B (thickness ~50 nm). The slope of line fit yields the rate of the strain modulation, $\alpha$. We obtain an $\alpha$ of $99 \pm 4$ meV/% in the armchair direction and $109 \pm 2$ meV/% in the zigzag direction. Data at $\varepsilon > 0.5\%$, where slippage may cause $E_g(\varepsilon)$ to artificially deviate from linear behaviour, are excluded from the line fit.

**Figure 3 | Room-temperature transport properties of black phosphorus FET under variable strain. a** and **b,** Ambipolar transport behavior of a black phosphorus FET under variable strain induced in armchair (**a**) and zigzag (**b**) direction. The CNP (marked by vertical broken lines) shifts slightly due to slight sample degradation between the two sets of measurements. The sample degradation is also responsible for the decrease of the peak value of $R$ in **b**. **c,** Conductance as a function of gate voltage under maximum compressive and tensile strain. Curves with strain induced in the armchair direction are shifted by 0.1 mS for clarity. Inset: Lack of strain dependence in the field-effect mobility. **d**, Piezo-resistive response at the CNP plotted as a function of strain. GF is obtained as the slope of line fits. $\mathrm{GF} = 122 \pm 10$ and $124 \pm 11$ for strains in the armchair and zigzag direction, respectively.

**Figure 4 | Black phosphorus strain sensor. a** and **b,** Piezo-resistive response as a function of strain measured at various gate voltages. Data sets with strain induced in the zigzag and armchair direction are plotted in **a** and **b**, respectively. The slope of line fits yields the GF at various gate voltages. **c,** GF plotted as a function of gate voltage for strains induced in both armchair and zigzag directions. GF reaches maximum at the CNPs (marked by vertical broken lines), whose position coincides with those in Fig. 3a

and 3b. **d** and **e,** Two-terminal resistance under repeated tensile (**d**) and compressive (**e**) strain manually induced in the armchair direction. The estimated GF is ~10. $V_g$ is kept at zero during measurement.

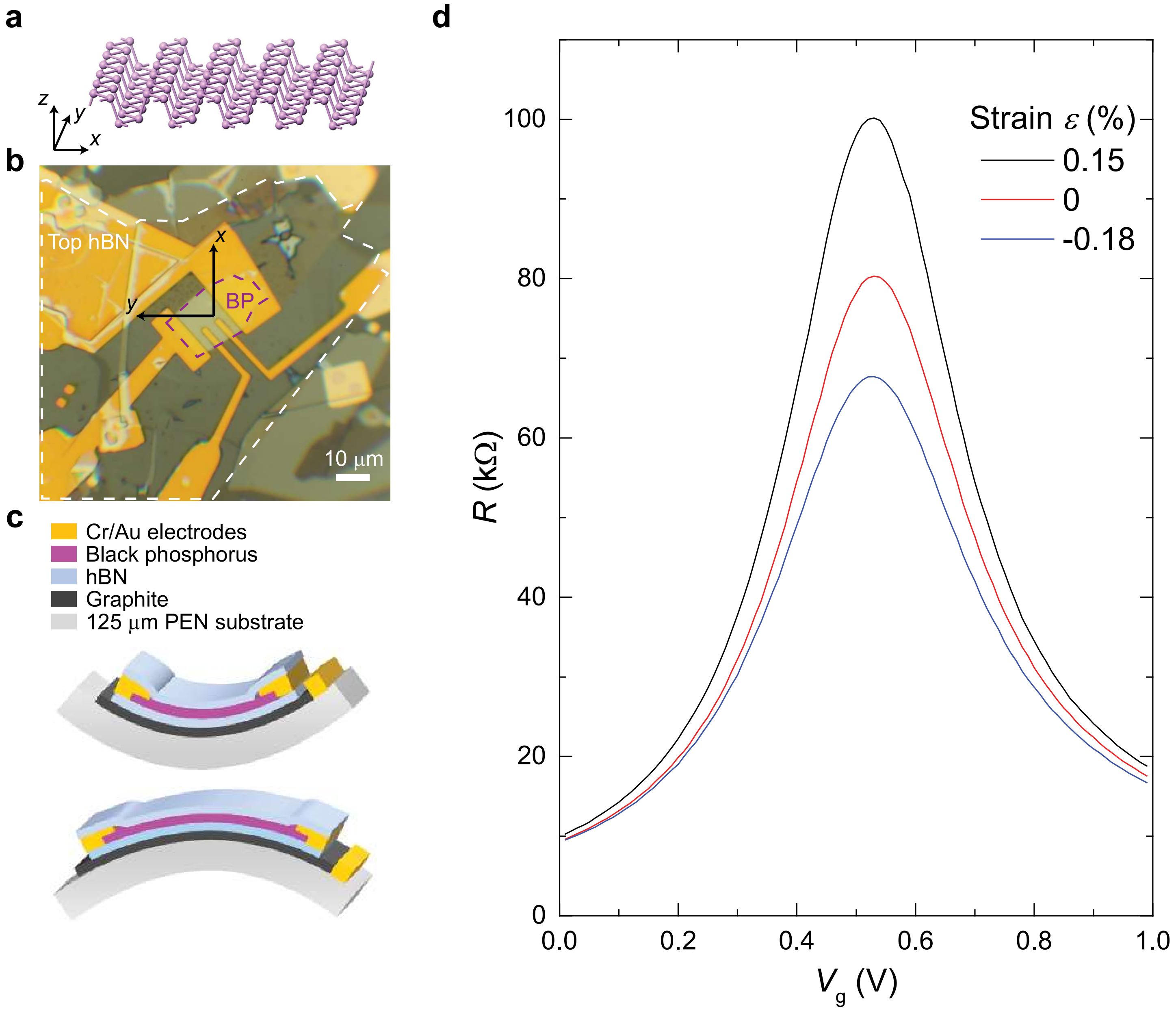

Figure 1

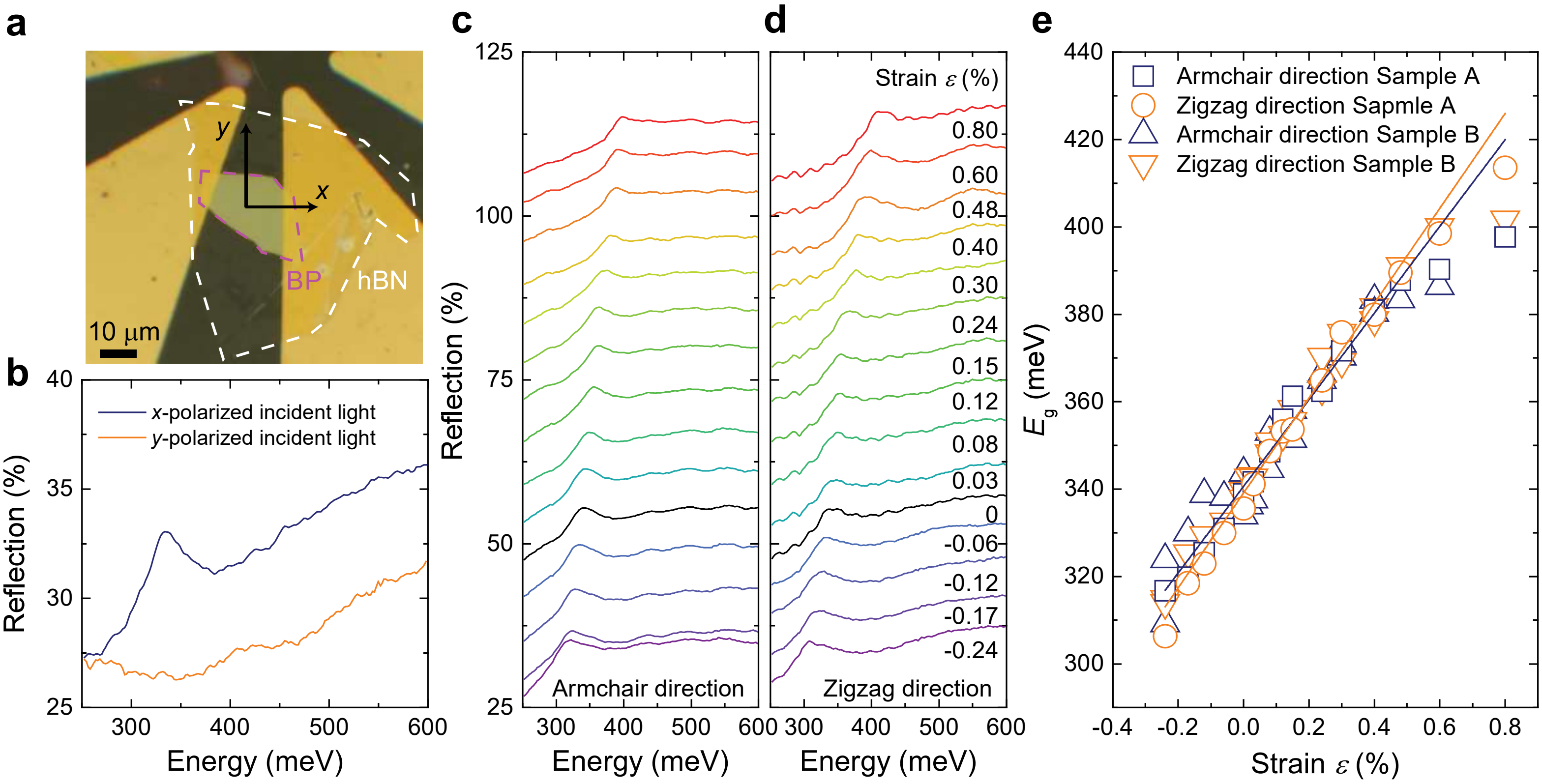

Figure 2

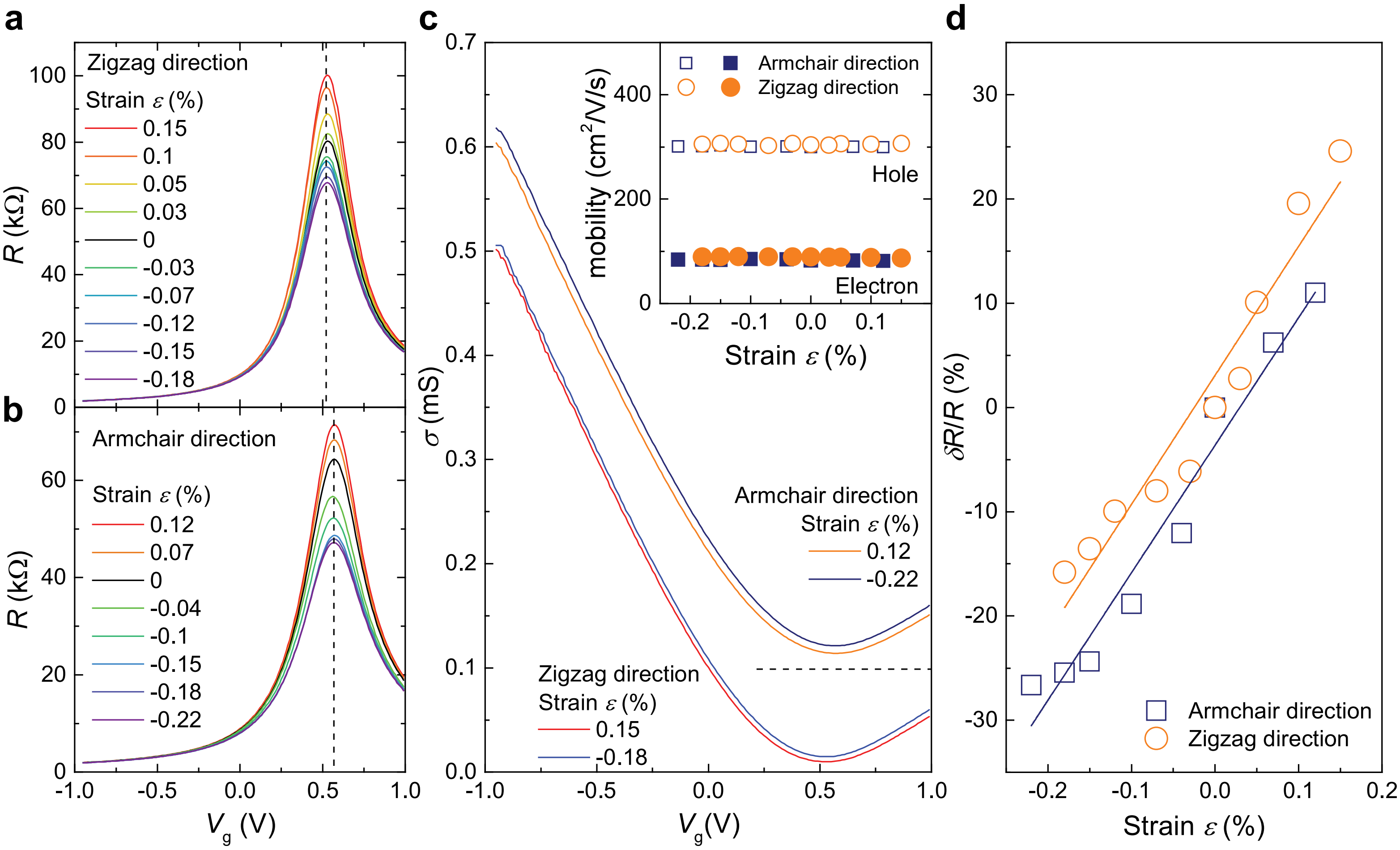

Figure 3

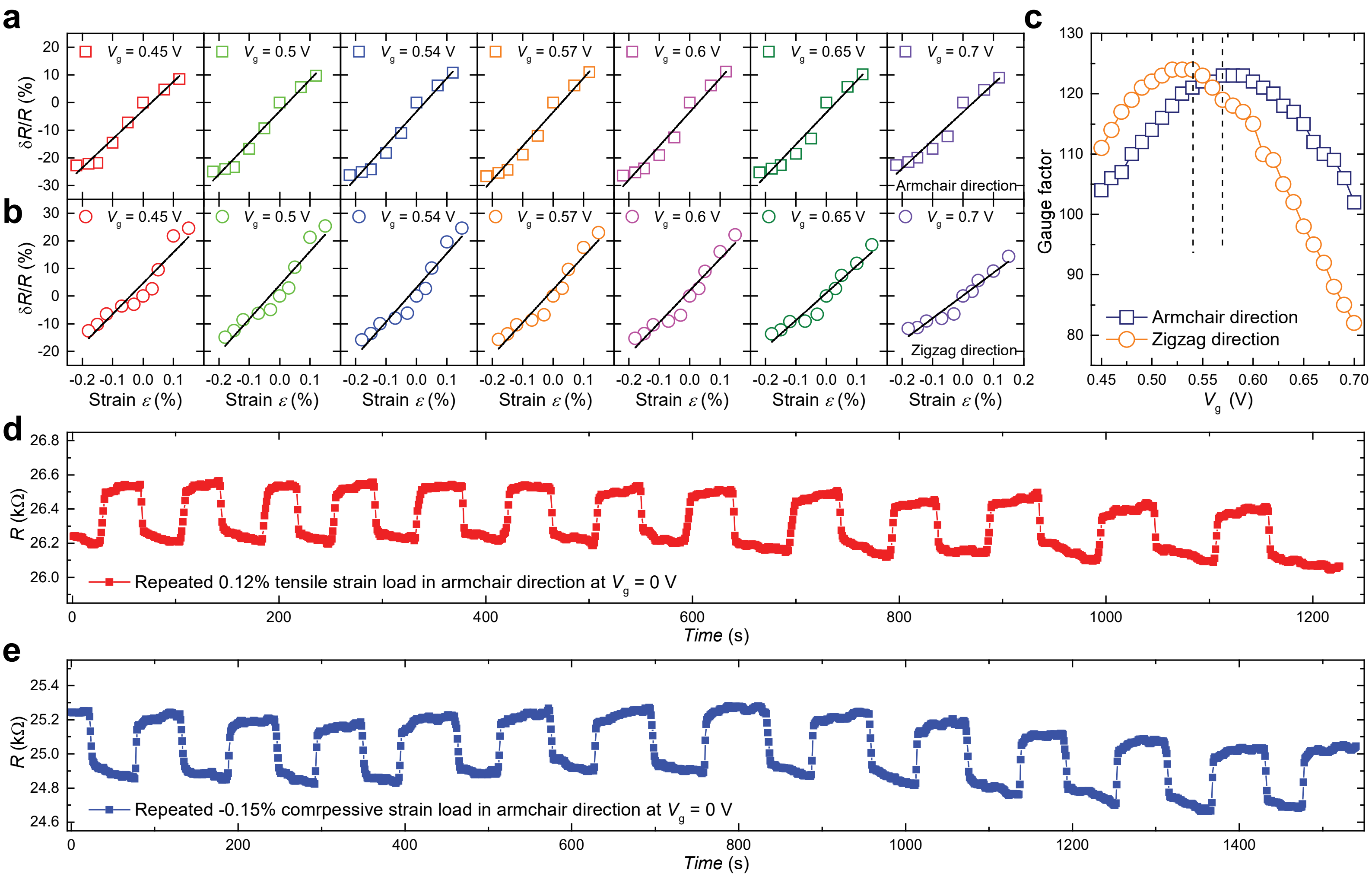
a
b
c
d
e
Vg = 0.45 V
Vg = 0.5 V
Vg = 0.54 V
Vg = 0.57 V
Vg = 0.6 V
Vg = 0.65 V
Vg = 0.7 V
δR/R (%)
Strain ε (%)
Armchair direction
Zigzag direction
Gauge factor
Vg (V)
R (kΩ)
Time (s)
Repeated 0.12% tensile strain load in armchair direction at Vg = 0 V
Repeated -0.15% comrpessive strain load in armchair direction at Vg = 0 V

Figure 4